\documentclass[aps,pra,longbibliography,amsmath,amssymb,reprint,superscriptaddress,floatfix]{revtex4-2}
\usepackage{graphicx}% Include figure files
\usepackage{amsmath}
\usepackage{amssymb}
\usepackage{float}
\usepackage{bm}% bold math
\usepackage[T1]{fontenc}
\usepackage[latin9]{inputenc}
\usepackage{nicefrac, xfrac}
\usepackage{wrapfig}
\usepackage{sidecap}
\usepackage{physics}
\usepackage{braket}
\usepackage{tikz}

\usepackage[unicode=true,pdfusetitle,
 bookmarks=true,bookmarksnumbered=false,bookmarksopen=false,
 breaklinks=false,pdfborder={0 0 1},backref=false,colorlinks=false]
 {hyperref}
\hypersetup{
 colorlinks,linkcolor=red,anchorcolor=black,citecolor=blue,breaklinks=true}
\makeatletter
\@ifundefined{textcolor}{}
{%
 \definecolor{BLACK}{gray}{0}
 \definecolor{WHITE}{gray}{1}
 \definecolor{RED}{rgb}{1,0,0}
 \definecolor{GREEN}{rgb}{0,1,0}
 \definecolor{BLUE}{rgb}{0,0,1}
 \definecolor{CYAN}{cmyk}{1,0,0,0}
 \definecolor{MAGENTA}{cmyk}{0,1,0,0}
 \definecolor{YELLOW}{cmyk}{0,0,1,0}
}
\makeatother

\allowdisplaybreaks

\begin{document}

\title{Merging Dipolar Supersolids in a Double-Well Potential}% Force line breaks with \\

\author{Hui Li}
\affiliation{JILA, NIST and Department of Physics, University of Colorado, Boulder, Colorado 80309, USA}
\email{hui.li@jila.colorado.edu}
\author{Eli Halperin}
\affiliation{JILA, NIST and Department of Physics, University of Colorado, Boulder, Colorado 80309, USA}
\author{Shai Ronen}
\affiliation{Cleerly Inc., 110 16$th$ Street, Suite 1400, Denver, CO, 80218, USA}
\author{John L. Bohn}
\affiliation{JILA, NIST and Department of Physics, University of Colorado, Boulder, Colorado 80309, USA}


\date{\today}% It is always \today, today,
\begin{abstract}
    We theoretically investigate the merging behaviour of two identical supersolids through dipolar Bose-Einstein condensates confined within a double-well potential. By adiabatically tuning the barrier height and the spacing between the two wells for specific trap aspect ratios, the two supersolids move toward each other and  lead to the emergence of a variety of ground state phases, including a supersolid state, a macrodroplet state, a ring state, and a labyrinth state. We construct a phase diagram that characterizes various states seen during the merging transition. Further, we calculate the force required to pull the two portions of the gas apart, finding that the merged supersolids act like a deformable plastic material.   Our work paves the way for future studies of layer structure in dipolar supersolids and the interaction between them in experiments. 
\end{abstract}

\maketitle

%\tableofcontents
\section{Introduction}

The formation of exotic quantum phases of matter unveils rich and fascinating 
 phenomena \cite{deguchi2012quantum, cinti2017superfluid, bottcher2020new, tanzi2021evidence, defenu2023long}.  In particular, the emergence of supersolidity has garnered significant attention \cite{gross1957unified, andreev1969quantum, chester1970speculations, leggett1970can, boninsegni2012colloquium, yukalov2020saga}. Supersolids exhibit both spatial periodicity and frictionless flow, which are usually incompatible in classical systems. Initially,  supersolidity was proposed to arise in solid $^4$He, however, despite some controversial claims, no conclusive evidence of supersolidity has been found in helium so far \cite{balibar2010enigma, kim2012absence}. 
 
 Alternatively, ultracold atomic gases have become a  platform for realizing and studying supersolids \cite{li2017stripe, leonard2017supersolid, bersano2019experimental, putra2020spatial}. In particular, dipolar Bose-Einstein condensates (BECs), composed of atoms with large magnetic or electric dipole moments, offer a unique opportunity to engineer the interatomic interactions which are crucial for supersolid formation. Dipolar interactions are long-ranged and anisotropic, factors which can induce density modulations and roton excitations in the BECs. Moreover, dipolar interactions can be tuned by applying external fields or by changing the geometry of the system \cite{bloch2008many, lahaye2009physics, chin2010feshbach, baranov2012condensed, bottcher2020new, chomaz2023dipolar}. 
 
 Dipolar BECs have been achieved in highly magnetic atoms, such as chromium \cite{griesmaier2005bose}, europium \cite{miyazawa2022bose}, dysprosium \cite{lu2011strongly}, and erbium \cite{aikawa2012bose}. Recent experiments have observed the appearance of a supersolid phase in elongated one-dimensional (1D) \cite{tanzi2019observation, bottcher2019transient, chomaz2019long, sohmen2021birth} and 2D \cite{norcia2021two, bland2022two} geometries with weakly magnetic dipole-dipole interactions (DDIs) in these dipolar BECs.

Additionally, ultracold bosons confined in a bilayer have drawn much attention because it is possible to control the inter- and intra-layer couplings \cite{gonzalez2019cold}. For example, a BEC loaded into spin-dependent optical lattices, forming a moir\'e structure, is the ideal scenario to investigate the physics behind superconductivity in twisted-bilayer graphene \cite{meng2023atomic}. Besides, dipolar gases in a layer-structure have been achieved with the help of an optical lattice potential \cite{koch2008stabilization, natale2022bloch, tobias2022reactions, du2023atomic} and stack supersolid structures have been discussed theoretically recently for an antidipolar single-component condensate \cite{mukherjee2023supersolid}, mixture of antidipolar and nondipolar condenstaes \cite{kirkby2023spin} and a doubly dipolar condensate with both electric and magnetic dipole moments \cite{ghosh2022droplet} confined in a standard harmonic trap.

 Double-well potentials serve as a versatile platform in the study of microscopic media, especially for atomic BECs, such as the quantum dynamics of ultracold atoms  \cite{smerzi1997quantum, spagnolli2017crossing, ferioli2019collisions, roy2022quantum}, self-trapping \cite{xiong2009symmetry, adhikari2009self, adhikari2014self}, stabilization of purely dipolar BEC and formation of quasi 2D sheets \cite{asad2009aligned}, and spin-squeezing \cite{tan2016spin}. A typical double well potential has been produced in experiments by adiabatic radio frequency-induced splitting~\cite{schumm2005matter, albiez2005direct, jo2007phase, leblanc2011dynamics} to study the dynamics of a splitting or merging process of BECs. 
 
 In this work, we combine concepts from dipolar BEC and double-well potentials. We propose applying a double well potential in the polarization direction of the atoms, then systematically bringing the two wells together while reducing the barrier to merge two supersolids. Along the way, many intermediate morphologies are seen.  

The paper is organized as follows. In Sec.~\ref{sec:th}, we outline the beyond mean-field theory and the numerical strategy and  introduce our system. Sec.~\ref{sec:lz2}  investigates the merging process in two cases, distinguished by the supersolids in the separated-well limit. Subsection~\ref{sec:lz2gs} considers supersolids in a trap with aspect ratio $\omega_z=2\omega$ prior to merging, while subsection~\ref{sec:lz3} considers a slightly different case, with initial aspect ratio $\omega_z=3\omega$. In both cases, the various morphologies are presented and summarized in phase diagrams in subsections~\ref{sec:lz2pd} and~\ref{sec:lz3}, respectively. A summary along with the future perspectives are given in Sec.~\ref{sec:co}.

\section{Theory model}
\label{sec:th}
\subsection{Formalism}
Here we consider a dipolar BEC composed of $N$ atoms of mass $m$, confined in an external potential to be defined below.  The ground-state wave function $\psi(\mathbf{r})$ satisfies the extended Gross-Pitaevskii equation (eGPE), which includes the effects of beyond-mean-field quantum fluctuations \cite{wachtler2016quantum, baillie2016self, saito2016path, roccuzzo2019supersolid}. The energy functional (energy per atom) of eGPE can be expressed as,
\begin{align}
    E(\psi) =& \int d\mathbf{r} \Big [ \frac{\hbar^2}{2m} |\nabla \psi(\mathbf{r})|^2 + V(\mathbf{r}) |\psi(\mathbf{r})|^2 +  \frac{gN}{2} \left | \psi(\mathbf{r}) \right |^4 \nonumber \\
    &+ \frac{N}{2} |\psi(\mathbf{r})|^2 \int d\mathbf{r}' U_{dd}(\mathbf{r} - \mathbf{r}') |\psi(\mathbf{r}')|^2 \nonumber \\
    &+ \frac{2N^{3/2}}{5} \gamma_\mathrm{QF}|\psi(\mathbf{r})|^5  \Big ],
    \label{eq:ef}
\end{align}
where $\hbar$ is the reduced Planck constant, the wave function is normalised as $\int |\psi(\mathbf{r})|^2 d\mathbf{r} =1$. The short-range two-body interaction coupling constant $g = 4\pi\hbar^2 a_s/m$ is governed by the $s$-wave scattering length $a_s$, and $U_{dd}(\mathbf{R})$ accounts for the anistropic and long-range DDIs given by
\begin{align}
    U_{dd}(\mathbf{R}) = \frac{3\hbar^2 a_{\mathrm{dd}}}{m} \frac{1-3\mathrm{cos}^2\theta}{|\mathbf{r}-\mathbf{r}'|^3},
\end{align}
with the dipole length $a_{\mathrm{dd}} = \mu_0 \mu_m^2 m /12\pi \hbar^2$, $\mu_0$ is the vacuum permeability and $\mu_m$ is magnetic moment, $\mathbf{R} \equiv \mathbf{r} - \mathbf{r}'$, we take $\hat{z}$ as the polarization axis, then $\theta$ denotes the angle between $\mathbf{R}$ and $\hat{z}$. The coefficient $\gamma_\mathrm{QF}$ represents the dipolar Lee-Huang-Yang correction within the local density approximation  \cite{lee1957many, lee1957eigenvalues, schutzhold2006mean, lima2011quantum, bisset2016ground, bottcher2020new}, induced by quantum fluctuation,
\begin{align}
   \gamma_\mathrm{QF} = \frac{32}{3}g\sqrt{\frac{a_s^3}{\pi}} \left (1+\frac{3}{2}\varepsilon_\mathrm{dd}^2 \right ),   
\end{align}
the dimensionless parameter $\varepsilon_\mathrm{dd} = a_\mathrm{dd} /a_s$. 

To study the merging process, we apply a double-well potential of the form
\begin{align}
   V(\mathbf{r}) = \frac{1}{2}m\omega^2(x^2 + y^2) + a(z^2 - z_0^2)^2,
   \label{eq:dw1}
\end{align}
with $z_0 >0$. $\omega$ is the harmonic trap frequency along the radial $x$ and $y$ directions. The parameters $a$ and $z_0$ control the central barrier height as well as the relative distance between the two wells \cite{zhang2012josephson, chen2022int, Pokatov_2023}.
When the two wells are separated, the double-well trap potential near its local minima $z = \pm z_0$ can be expanded with a harmonic frequency $\omega_z$, one can obtain a harmonic form $V_\mathrm{ho}(z)\doteq \frac{1}{2}m\omega_z^2 (z\pm z_\mathrm{0})^2$, with $m\omega_z^2 = 8az_0^2$, where $\omega_z$ corresponds to the trap frequency along the axial $z$ direction. We can rewrite the double-well potential with the help of $\omega_z$,
\begin{align}
    V(\mathbf{r}) = \frac{1}{2}m\omega^2(x^2 + y^2) + \frac{m\omega_z^2}{8z_0^2}(z^2 - z_0^2)^2,
    \label{eq:dw2}
\end{align}
where the barrier height is $V_h = m\omega_z^2z_0^2/8$. Thus the two wells each have the specified frequency $\omega_z$ when they are far from each other, $i.e.$, $z_0 \gg a_{ho}$  In this way one can specify the pre-merged condensates by the separated-trap aspect ratio $\lambda = \omega_z / \omega$.

\subsection{Numerical strategy}

Ground state wave functions can be obtained by numerically minimizing the energy functional Eq.~\ref{eq:ef}. 
We employ an optimization algorithm limited-memory Broyden-Fletcher-Goldfarb-Shanno (L-BFGS) method \cite{liu1989limited, zhu1997algorithm}, which has been implemented in PyTorch \cite{paszke2019pytorch}.
Compared to other first-order gradient descent algorithms, L-BFGS shows more stable and faster convergence, specifically, its superior performance in terms of computational time can be enhanced when running on a graphics processing unit (GPU). 

The condensate wave function is discretized with an equally spaced 3D grid on a cube box centered on the origin. Because the dipolar interaction integral diverges in configuration space, the integration has been calculated in momentum space using an analytical form of the interaction, with a spherical cut-off \cite{ronen2006bogoliubov}. We employed a cubic grid of $256\times 256\times 256$ grid points with dimensions $56\times 56 \times 56$ $a_\mathrm{ho}$, which ensures a large enough grid to allow the various terms in the energy functional to be computed with spectral accuracy.

In this work, we consider $^{162}$Dy in a trap with radial trap frequency $\omega = 2 \pi \times 125$ Hz, $a_s = 85 a_0$ and $a_\mathrm{dd} = 131 a_0$, here $a_0$ is the Bohr radius. 
%The trap aspect ratio is defined by $\lambda_z \equiv \omega_z/\omega$.
The harmonic-oscillator length $a_\mathrm{ho} = \sqrt{\hbar/m \omega}$ and an energy units of $\hbar \omega$ will be used to scale the length and energy. 

\begin{figure}[ht]
    \centering
    \includegraphics[width=0.95\columnwidth]{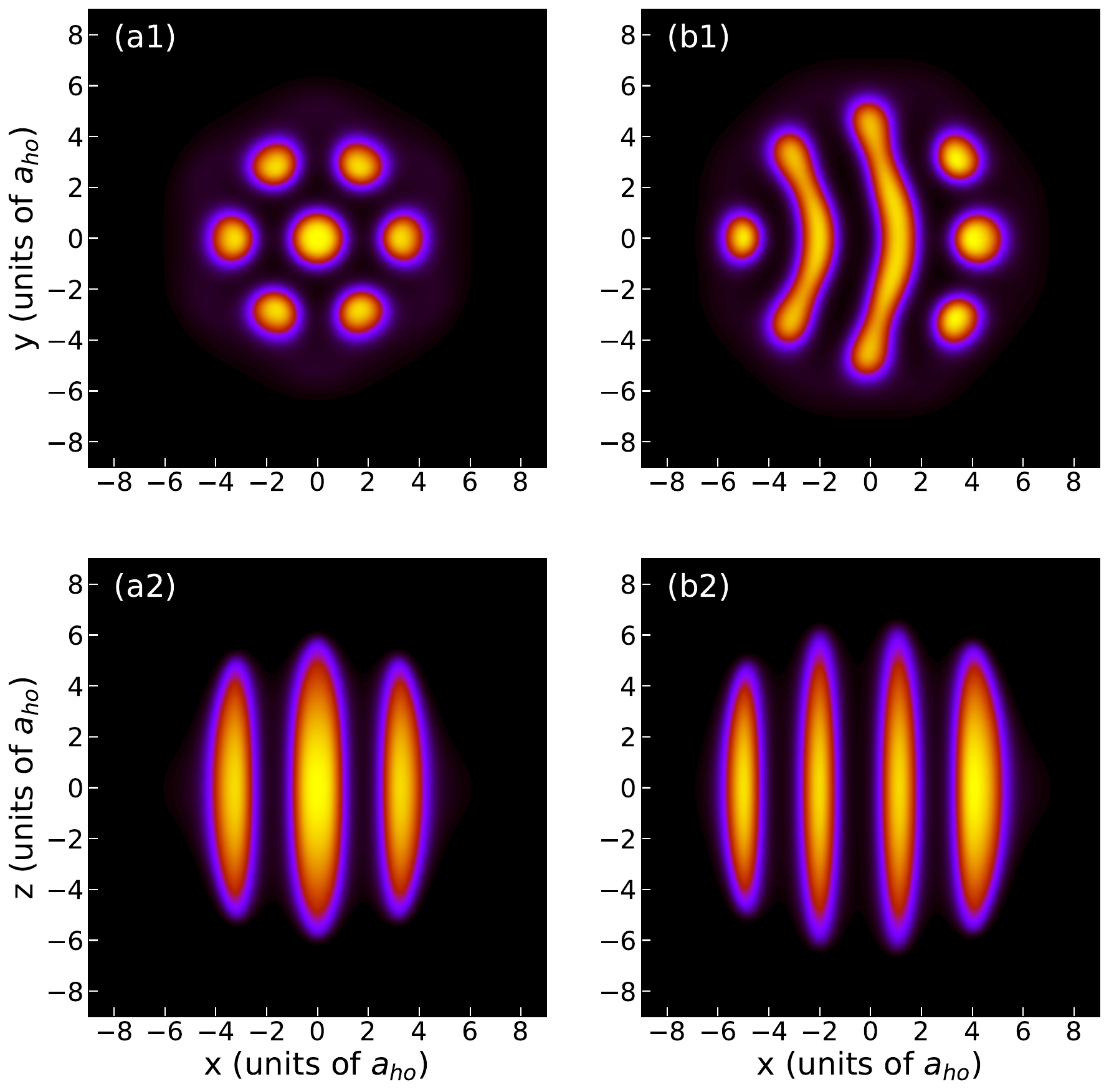}
    \caption{2D Slices of the density profiles for the dipolar BEC with atom numbers of $N=10^5$ (a1), (a2), and $N=2 \times10^5$ (b1), (b2) $^{162}$Dy confined in the harmonic trap with $\omega_z = 2\omega$.}
    \label{fig:1}
\end{figure}

\section{Numerical Results} % for \texorpdfstring{$\omega_z = 2\omega$}{b}}
\label{sec:lz2}

\subsection{Characterization of merging process}
\label{sec:lz2gs}

Our goal in this work is to monitor the behavior of two supersolids as they merge together.  To this end, we begin with the situation at the beginning and end of this process.  In this section we consider, in the separated limit, two ``standard'' supersolids such as the one shown in Figs.~\ref{fig:1}(a1) and (a2). This supersolid consists of $N=10^5$ atoms in a single harmonic trap with frequencies $\omega = 2 \pi \times 125$ Hz, and $\omega_z = 2 \omega$, defined by the trap potential 
\begin{align}
   V(\mathbf{r}) = \frac{1}{2}m\omega^2(x^2 + y^2 + 4x^2).
   \label{eq:sw}
\end{align}
 In these circumstances, dysprosium is well-known to display a six-fold symmetric supersolid phase~\cite{baillie2018droplet, poli2021maintaining, norcia2021two, bland2022two, young2022supersolid, halperin2023frustration}.

A single such supersolid in this trap is depicted in Figure \ref{fig:1}. Fig.~\ref{fig:1}(a1) shows a cut through the density at $z = 0$ of the 3D density $|\psi(x, y, 0)|^2$, and from the side at $y=0$ in Fig.~\ref{fig:1}(a2). The energy per atom is $E = 22 \ \hbar\omega$ and the chemical potential is $\mu_0 = 31 \ \hbar\omega$. In one limit of the simulation, we envision two such supersolids, centered at the vertical coordinates $\pm z_0$, far enough away that they only weakly perturb one another. 

\begin{figure}[ht]
    \centering
    \includegraphics[width=0.95\columnwidth]{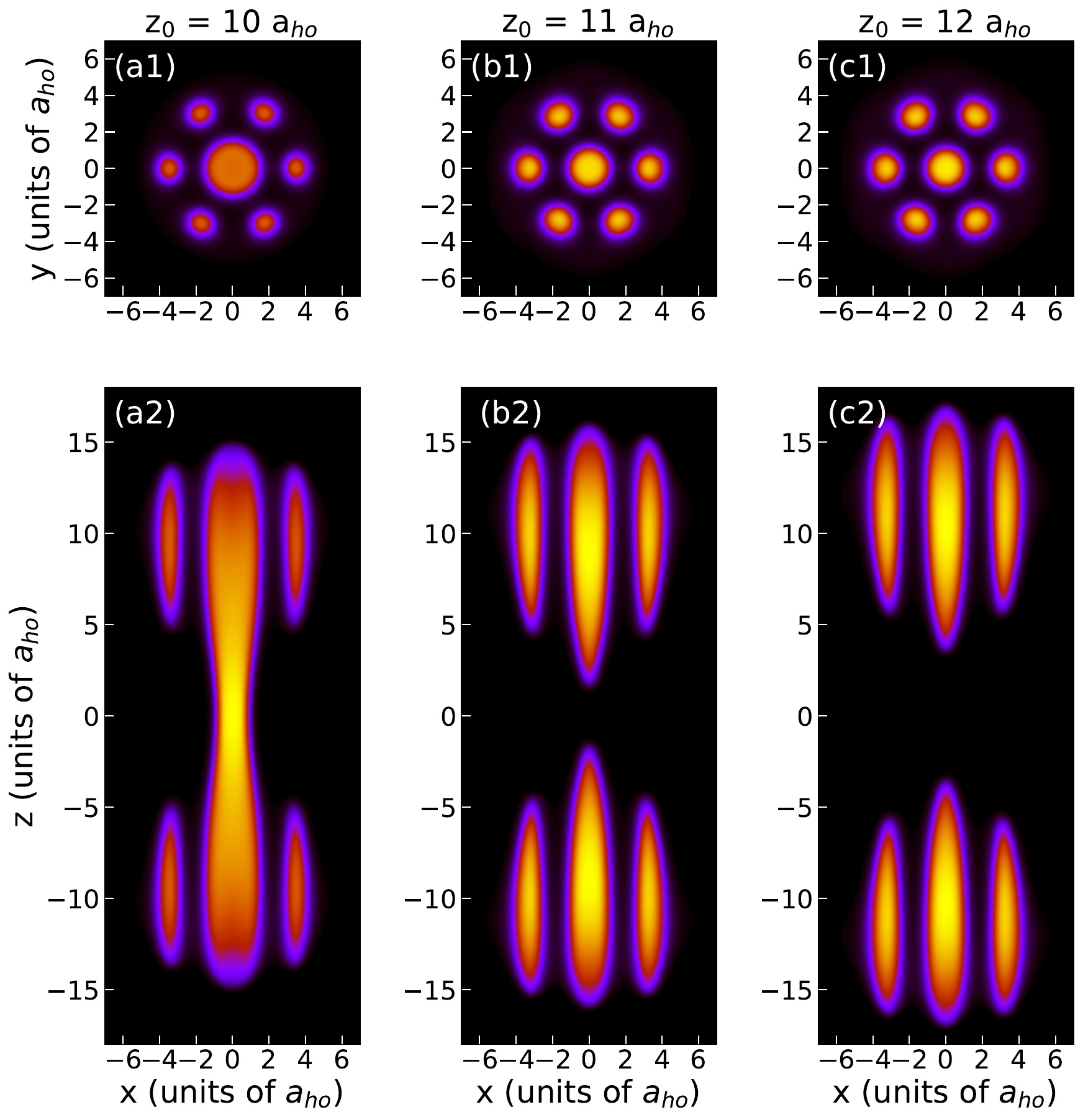}
    \caption{2D Slices of the density profiles of the dipolar BEC of $N=2 \times 10^5$ $^{162}$Dy atoms confined in the double well trap with $\omega_z = 2\omega$. The upper panels show the cut of the $x-y$ plane at the local minima of $z$ with (a1) $z_0 = 10 \ a_\mathrm{ho}$, (b1) $z_0 = 11 \ a_\mathrm{ho}$, and (c1) $z_0 = 12 \ a_\mathrm{ho}$. The lower panels illustrate the corresponding cut along the $x-z$ plane at $y=0$.}
    \label{fig:2}
\end{figure}

In the merged limit where $z_0=0$, a different BEC configuration would be realized, as shown in Figs.~\ref{fig:1}(b1) and (b2). In this case the confining potential is a single harmonic trap as described by Eq.~\ref{eq:sw}, with familiar characteristics $\omega = 2 \pi \times 125$ Hz, $\omega_z = 2 \omega$,  whereas the merged BEC now contains $N = 2 \times 10^5$ atoms, and is therefore in a different part of the supersolid phase space. In this case it is more like a labyrinth, as demonstrated in Ref.~\cite{hertkorn2021pattern}, higher atom numbers at the same scattering length can break up the supersolid droplets into labyrinth phase. Figures \ref{fig:1}(b1) and (b2) therefore represent the anticipated result of fully merging the two supersolids into a single layer.

\begin{figure}[ht]
    \centering
    \includegraphics[width=0.95\columnwidth]{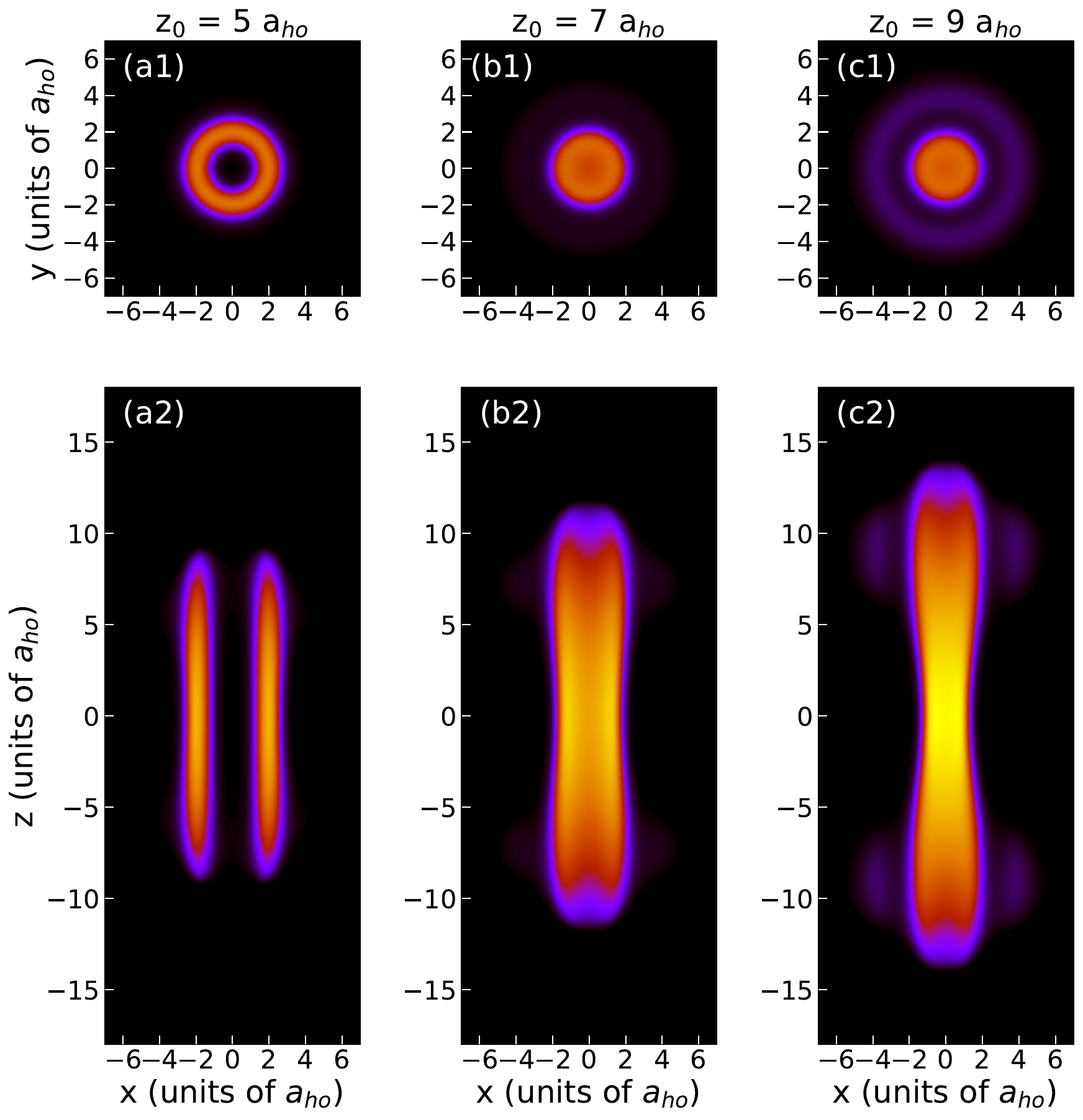}
    \caption{2D Slices of the density profiles of the dipolar BEC of $N=2 \times 10^5$ $^{162}$Dy atoms confined in the double well trap with $\omega_z = 2\omega$. The upper panels show the cut of the $x-y$ plane at the local minima of $z$ with (a1) $z_0 = 5 \ a_\mathrm{ho}$, (b1) $z_0 = 7 \ a_\mathrm{ho}$, and (c1) $z_0 = 9 \ a_\mathrm{ho}$. The lower panels illustrate the corresponding cut along the $x-z$ plane at $y=0$.}
    \label{fig:3}
\end{figure}

The initial stages of the merger, for relatively large $z_0$, are shown in Fig.~\ref{fig:2}.  In these panels the values of $z_0$ are, from left to right, $z_0 = 10,11,12 \ a_\mathrm{ho}$. Shown are side view cuts through $y=0$ (lower panels) and top view cuts through $z=z_0$ (upper panels). The double well potential is given by Eqs.~\ref{eq:dw1} and \ref{eq:dw2}.

\begin{figure}[ht]
    \centering
    \includegraphics[width=1\columnwidth]{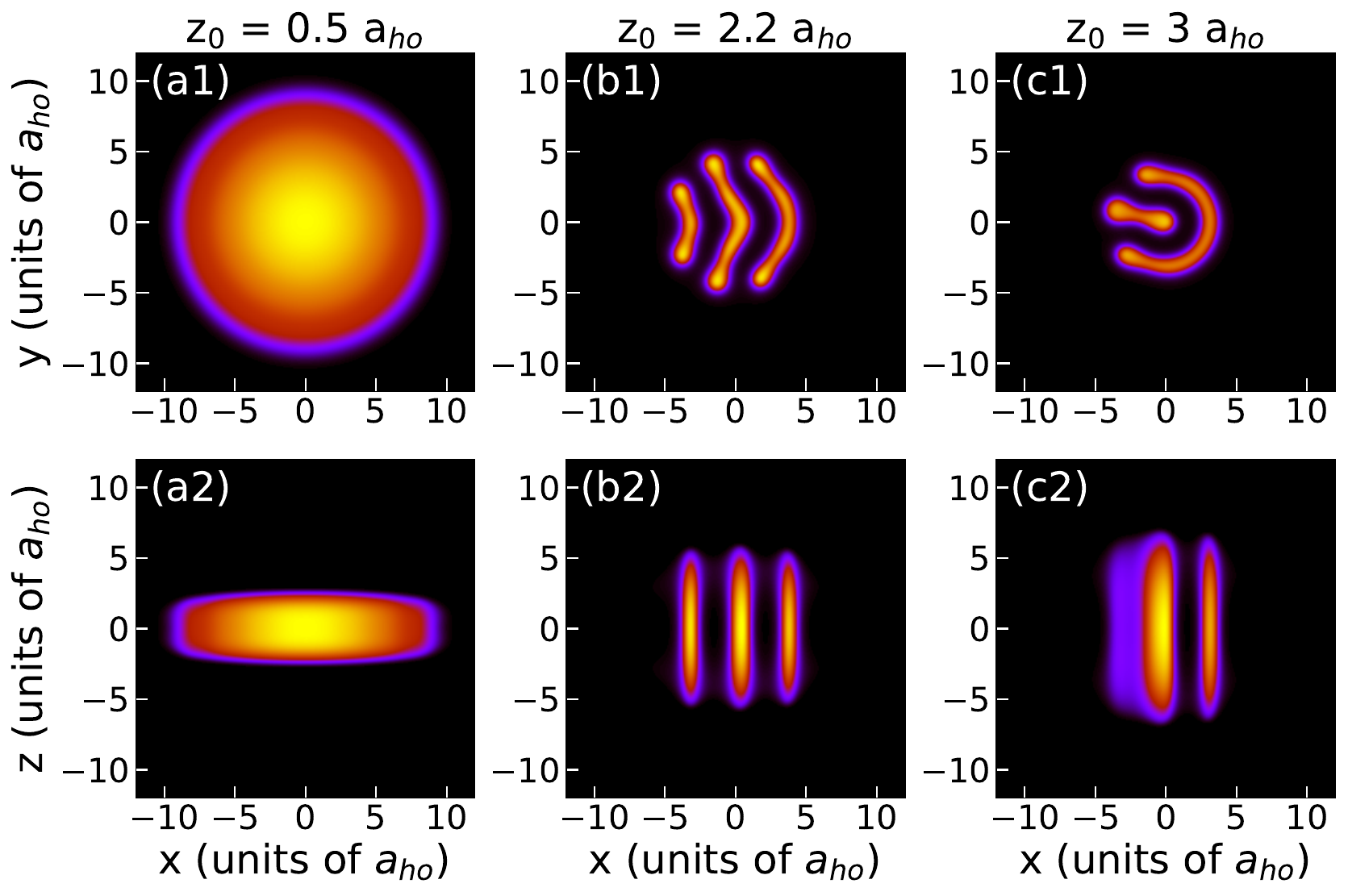}
    \caption{2D Slices of the density profiles of the dipolar BEC of $N=2 \times 10^5$ $^{162}$Dy atoms confined in the double well trap with $\omega_z = 2\omega$. The upper panels show the cut of the $x-y$ plane at the local minima of $z$ with (a1) $z_0 = 0.5 \ a_\mathrm{ho}$, (b1) $z_0 = 2.2 \ a_\mathrm{ho}$, and (c1) $z_0 = 3 \ a_\mathrm{ho}$. The lower panels illustrate the corresponding cut along the $x-z$ plane at $y=0$.}
    \label{fig:4}
\end{figure}

At the largest separation $z_0=12 \ a_\mathrm{ho}$ [e.g., Figs.~\ref{fig:2}(c1) and (c2)], the gas is still fundamentally separated into two supersolids. Their long-range attraction to each other is evident in the way the central density maximum in each is drawn toward the center at $z=0$.  Between $z_0=11 \ a_\mathrm{ho}$ [e.g., Figs.~\ref{fig:2}(b1) and (b2)] and $z_0=10 \ a_\mathrm{ho}$ [e.g., Figs.~\ref{fig:2}(a1) and (a2)] this attraction becomes overwhelming, and the BEC easily tunnels through the barrier; this is the occasion of merging.  Once this pathway is open between the upper and lower components, atomic density floods into this central tendril, reducing the density in the six local density maxima on the periphery.  

The process of merging continues in Fig.~\ref{fig:3}, which shows similar density plots for (from left to right) $z_0 = 5, 7, 9 \ a_\mathrm{ho}$. By $z_0=9 \ a_\mathrm{ho}$, the original six density minima around the periphery of the supersolid have all but vanished, leaving an elongated cloud that thins in the middle as if it were a stretched piece of taffy.  Upon further merging to smaller $z_0$, tunneling becomes easier, the gas conforms to a more cylindrical shape, and ultimately  forms a cylindrical shell, a shape familiar from previous work \cite{poli2021maintaining}. At even smaller $z_0$ the labyrinth states begin to appear, two examples of which are illustrated in Figs.~\ref{fig:4}(b1), (b2), and (c1), (c2). Finally, the emergence of a pancake shaped BEC state [e.g., Figs.~\ref{fig:4}(a1) and (a2)] has been observed when $z_0$ close to 0.  

Note that in Fig.~\ref{fig:2}, the upper and lower supersolids are displayed radially so that the features above are aligned with the features below. This is done to simplify the figure; in fact, the DDI is sufficiently weak in this limit that no such alignment necessarily occurs.  We verified this by artificially rotating the upper supersolid relative to the lower one, finding the energy variation to be only the order of $10^{-5} \ \hbar \omega$.

\subsection{Phase diagram}
\label{sec:lz2pd}

\begin{figure}[ht]
    \centering
    \includegraphics[width=1\columnwidth]{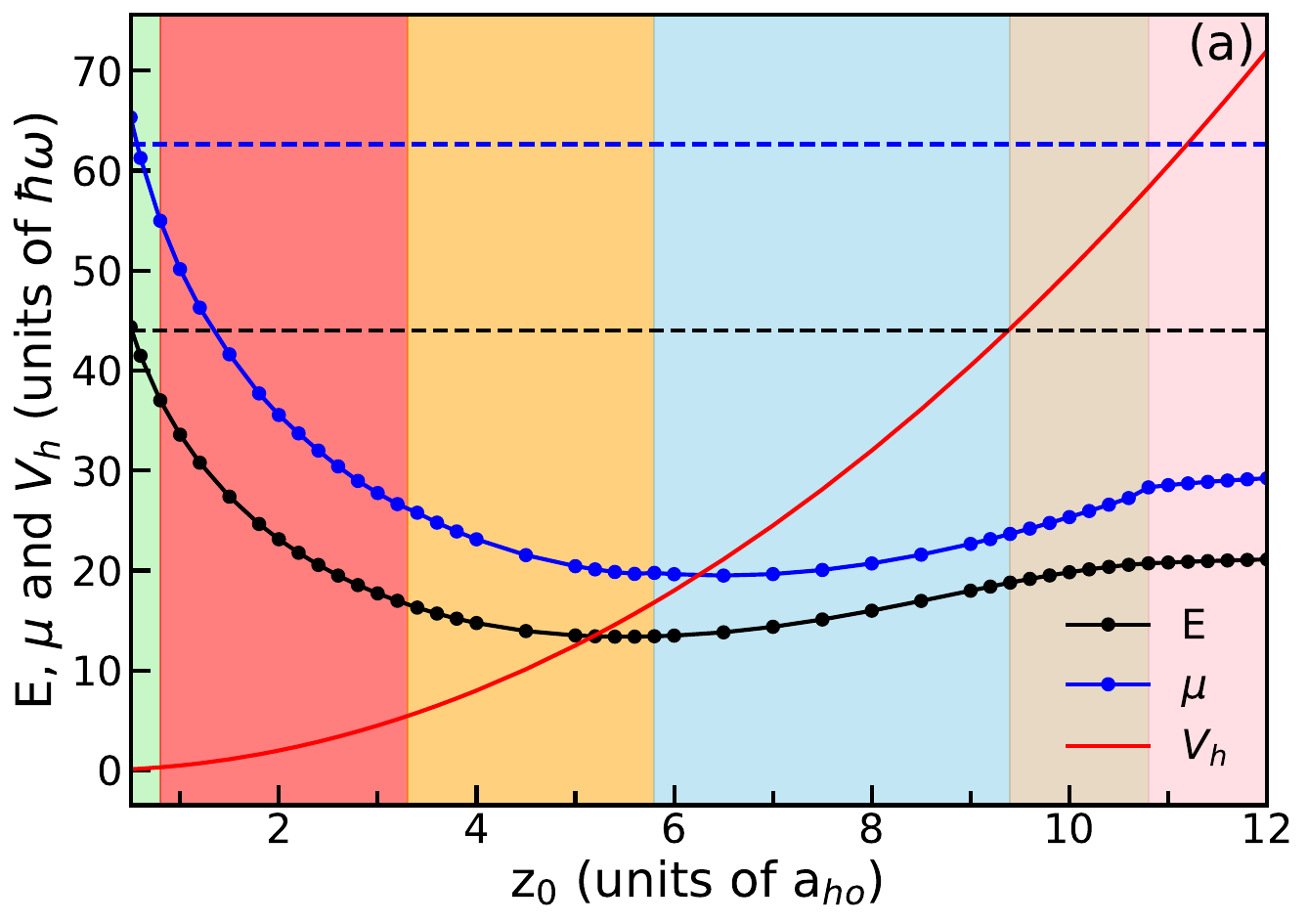}
    \includegraphics[width=1\columnwidth]{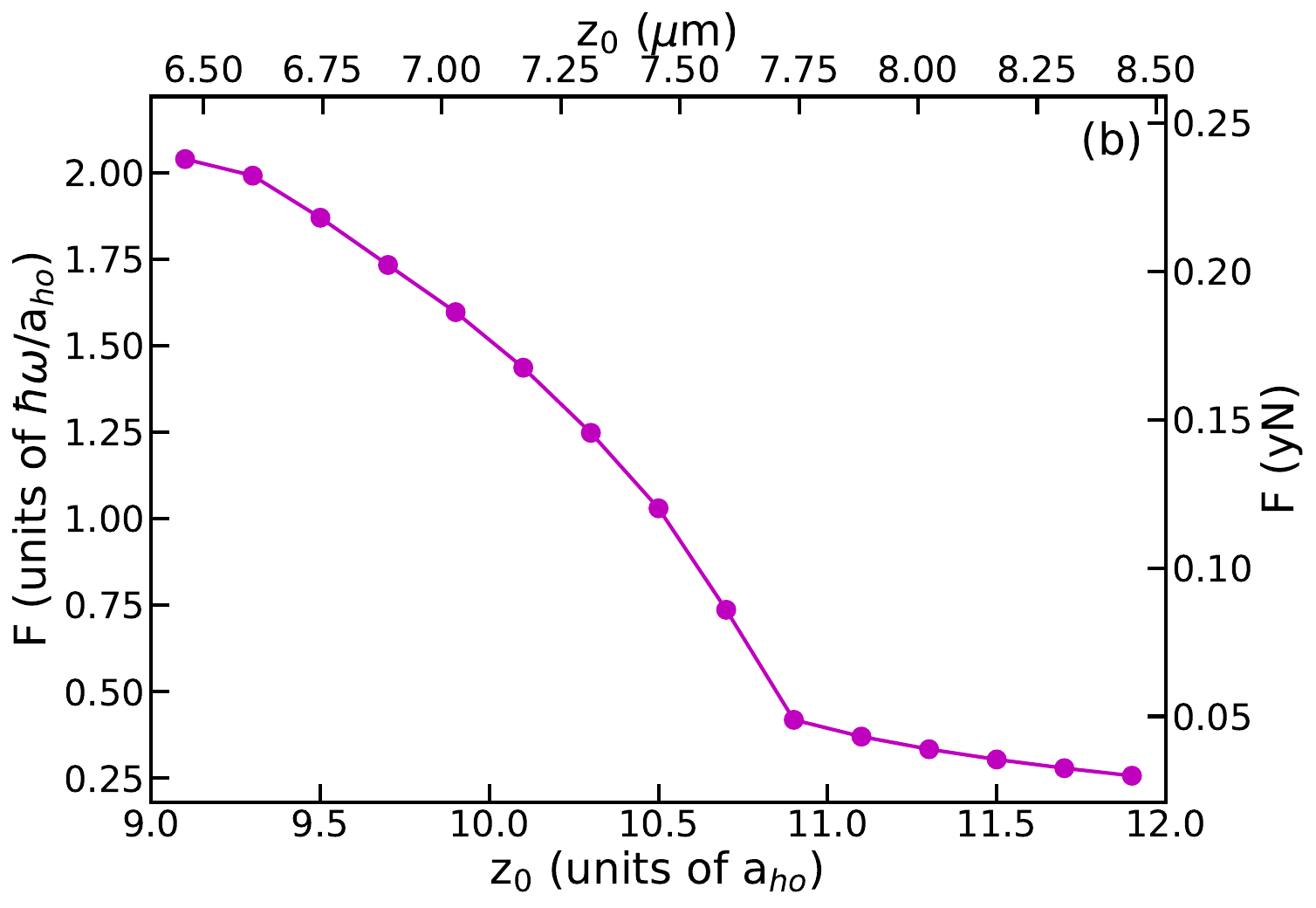}
    \caption{(a) Phase diagram of the dipolar BEC of $N=2 \times 10^5$ $^{162}$Dy atoms confined in the double well trap with $\omega_z = 2\omega$, the energy per atom (thick black curve) and chemical potential (thick blue curve) and the barrier height as a function of the local minima parameter $z_0$ are shown. The black and blue dashed lines are twice of energy per atom and chemical potential for $N= 10^5$ atoms confined in the single harmonic trap with the same trap aspect ratio. (b) The stretch force, explained in the context, as a function of $z_0$ around $z_0 = 11 \ a_\mathrm{ho}$. }
    \label{fig:5}
\end{figure}

Thus various morphologies appear during the merging process, as summarized by the colored regions in Fig.~\ref{fig:5}(a).  To assist in the description of what happens as the gas evolves between these morphologies, we also plot the energy per atom $E$ (black), the chemical potential $\mu$ (blue), and the height of the barrier separating the upper and lower wells (red).  

Perhaps the most important part of this diagram is around $z_0=11 \ a_\mathrm{ho}$, where the transition occurs between separated supersolids (pink region) and those that have merged (tan region), as shown in Figs.~\ref{fig:2}(a1), (a2) and (b1), (b2). For the separated supersolids, both the energy and chemical potential diminish very slightly as the two clouds are brought together, presumably due to the long-range dipole attraction between them.

\begin{figure*}[ht]
    \centering
    \includegraphics[scale=0.3,trim=10 5 15 0,clip]{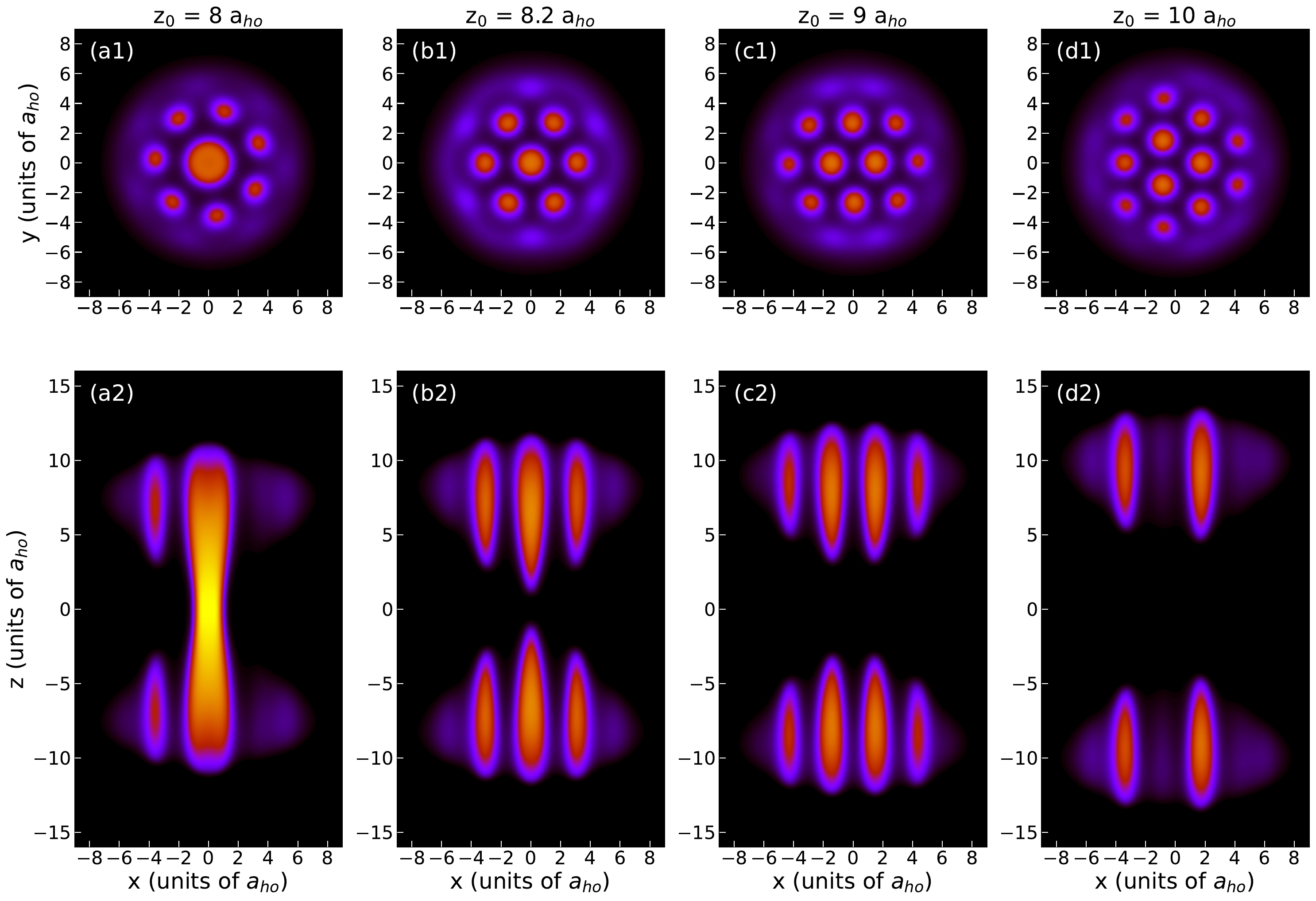}
    \caption{2D Slices of the density profiles of the dipolar BEC of $N=2 \times 10^5$ $^{162}$Dy atoms confined in the double well trap while holding fixed $\omega_z = 3\omega$. The upper panels show the cut of the $x-y$ plane at the local minima of $z$ with (a1) $z_0 = 8 \ a_\mathrm{ho}$, (b1) $z_0 = 8.2 \ a_\mathrm{ho}$, (c1) $z_0 = 9 \ a_\mathrm{ho}$, and (d1) $z_0 = 10 \ a_\mathrm{ho}$. The lower panels illustrate the corresponding cut along the $x-z$ plane at $y=0$.}
    \label{fig:7}
\end{figure*}

Once they merge, however, the energy and chemical potential drop much more rapidly as $z_0$ decreases, and the barrier height (red line) drops below twice the chemical potential of a single, isolated supersolid (blue dashed line). Here, a long, thin, continuous strip of dipoles aligned along the strip's axis defines a tendril that connects the lower and upper branches of the condensate.  Viewing the gradient of energy with respect to $z_0$ as a force, Fig.~\ref{fig:5}(b) plots the force required to pull the two condensates apart. This force decreases as the distance increases and the tendril becomes thinner, behavior reminiscent of stretching a plasticine material such as taffy. At the phase transition, the tendril breaks and the force has a different dependence on density. This dependence is due to the usual long-term dipolar attraction.

The next milestone occurs as a transition from the merged supersolids (tan region) to the single elongated object in Figs.~\ref{fig:3}(b1) and (b2) (skyblue region).  This is seen to occur when the barrier height (red line) drops below twice of the energy per atom of a single, isolated supersolid (black dashed line). At this point it is no longer energetically favorable for the gas to form the six density maxima at the periphery of the supersolid.  The hexagonal pattern associated with the supersolid is gone.

Next, the BEC passes into the cylindrical morphology of Figs~\ref{fig:3} (a1) and (a2) (orange region).  At this point the BEC, being shortened in the vertical direction, experiences increasing dipolar repulsion, thus the total energy and chemical potential now rise with decreasing $z_0$. This shift in the energy balance occasions the transformation to the hollow cylinder, in which the dipoles repel one another, in the same way that the electrons in an electrically charged conductor flow to the conductor's surface.

When $z_0$ is further reduced, the hollow cylinder breaks up their symmetry and enter the labyrinthine phase (red region), where various near degenerate states coexist in this regime as examples shown in Figs.~\ref{fig:4} (b1), (b2) and (c1), (c2) . Finally, in the $z_0 \rightarrow 0$ limit, the gas assumes a simple pancake-like shape as Figs.~\ref{fig:4} (a1) and (a2) (green region). Note that in this case, the confinement in the vertical direction is given by a nearly quartic, rather than quadratic, potential. Because of this difference, the result of merging is actually distinct from the labyrinth predicted in Figs. \ref{fig:1}(b1) and (b2).

\subsection{Increasing the axial trap frequency}
\label{sec:lz3}

Here, we turn to investigate the merging behavior by increasing the vertical confinement frequency to $\omega_z = 3 \omega$ with the same frequency of $\omega = 2 \pi \times 125$ Hz for the radial directions. In this situation, the trap aspect ratio is more oblate than in the previous section, which allows a more elaborate pattern to develop even in a single layer.  This is shown in Figs.~\ref{fig:a1}(a1) and (a2) which show the density profile of $N=1 \times 10^5$, in this case a 12-droplet hexagonal supersolid state, which has been predicted in Ref.~\cite{poli2021maintaining}. By contrast, in the the merged limit where $N=2\times10^5$ atoms are present,  a labyrinthine state is seen, in Figs.~\ref{fig:a1}(b1) and (b2). 

Upon merging the supersolids by reducing $z_0$, an even greater variety of morphologies is seen. Similar to the $\omega_z = 2\omega$ case, the gas is separated into two supersolids at large $z_0$ at first, as shown in Fig.~\ref{fig:7}. As $z_0$ decreases, the number of distinct droplets in the pattern drops from 12 at $z_0= 10 a_\mathrm{ho}$,  [Figs.~\ref{fig:7}(d1) and (d2)] transitions to 10 droplets at $z_0= 9 a_\mathrm{ho}$ [Figs.~\ref{fig:7}(c1) and (c2)], then to 7 droplets (with lesser features on the periphery) at $z_0= 8.2 a_\mathrm{ho}$ [Figs.~\ref{fig:7}(b1) and (b2)]. Finally, by $z_0= 8 a_\mathrm{ho}$ the upper and lower supersolids have merged [Figs.~\ref{fig:7}(a1) and (a2)], surrounded by seven additional features, as compared to the six features surrounding the central tendril in the $\omega_z = 2 \omega$ case.

\begin{figure}[ht]
    \centering
    \includegraphics[width=1\columnwidth]{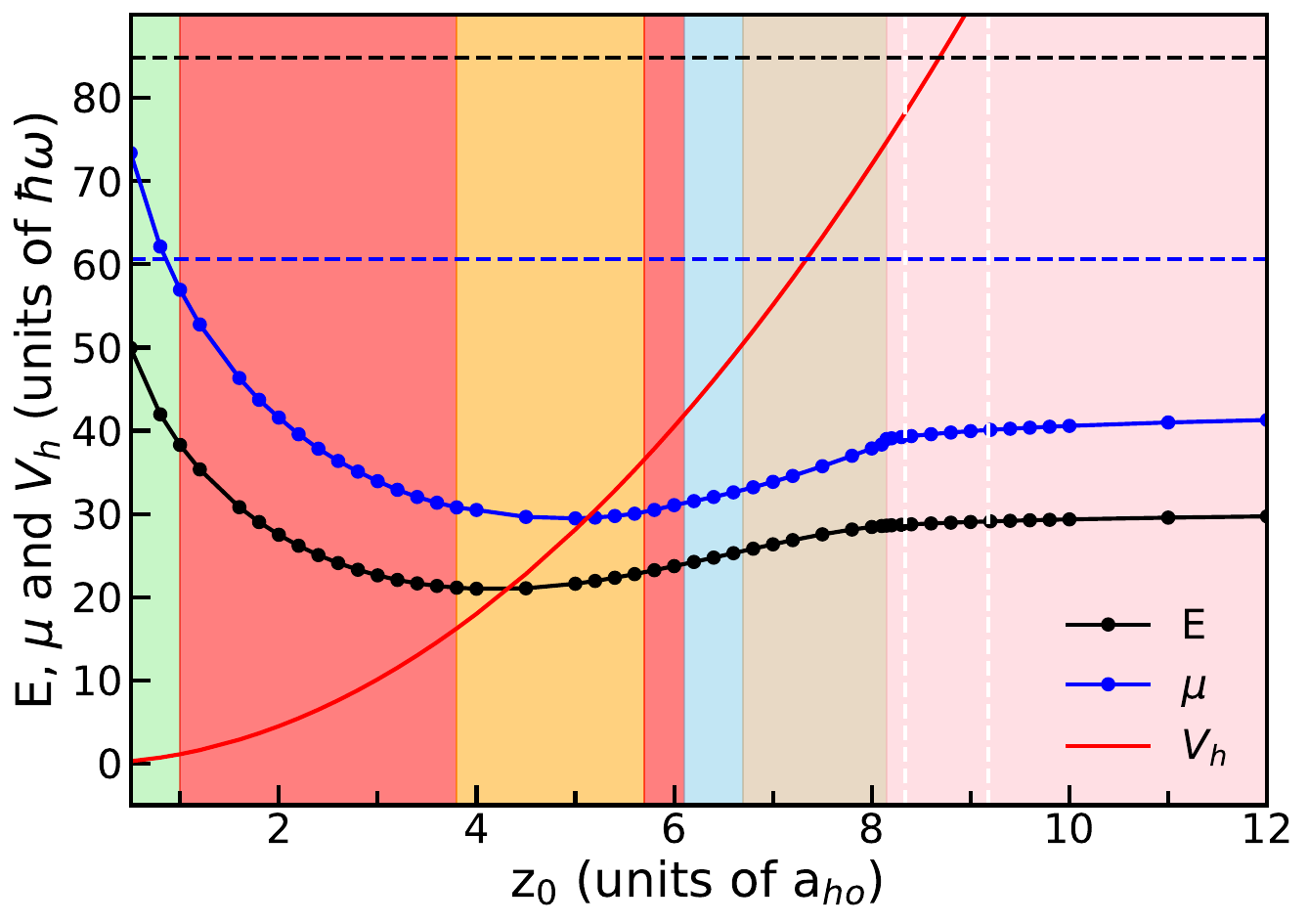}
    \caption{Phase diagram of the dipolar BEC of $N=2 \times 10^5$ $^{162}$Dy atoms confined in the double well trap with $\omega_z = 3\omega$. The chemical potential (thick blue curve) and the energy per atom (thick black curve) and the barrier height (red) as a function of the local minima parameter $z_0$. The blue and black dashed lines are twice of the chemical potential and energy per atom for $N= 10^5$ atoms confined in the harmonic traps. The two white dashed lines divide the 7-, 10-, and 12-droplets regimes from left to right.}
    \label{fig:10}
\end{figure}

At intermediate values of $z_0$, the cylindrical form of the connecting tendril appears, similar to what was observed in the case with $\omega_z = 2\omega$ above [Fig.~\ref{fig:8}]. Further reducing $z_0$ results in a small window of labyrinthine patterns before entering the cylindrical shell regime, and then the other labyrinthine. Finally, in the merged limit,  $z_0 \le 2.5$, concentric rings appear, the outermost distributed in a large number of droplets [eg, Figs.~\ref{fig:9}(b1) and (b2)]. A phase diagram articulating these morphologies is shown in Fig.~\ref{fig:10}. The different colors indicate the similar phase state region as described in Sec.~\ref{sec:lz2pd}.

\section{Conclusions and outlooks}
\label{sec:co}
In addition to the vast array of patterns known to occur for individual dipolar BECs in pancake-shaped traps, we have now demonstrated that even a greater variety exists when two such BECs are allowed to influence one another. This influence already occurs at a distance when the clouds are relatively far apart, but results in a greater variety after they have merged.

We have further shown that, close to the merging transition, the BECs are connected by a thin tendril of dipolar material with an internal cohesion.  This tendril behaves much like a deformable, plasticine material, whose cohesive force diminishes as the tendril lengthens and thins. The merging transition is characterized by the abrupt change by a discontinuity in the derivative of this minute force, as the tendril breaks.

The ground-state phase diagram have been identified, and typical density profile cutoffs are showcased. Building upon these results, the quench dynamics and the collective Bogoliubov excitation spectrum across the phase transition will be fascinating for better understanding the merge process. With the help of a double well potential, it is also interesting to exploit the collision process of two supersolids. Besides, The similar behaviour of two-components or antidipolar single-component supersolids can be a promising research directions in both theory and experiment. 

\section*{Acknowledgements}
%\nocite{*}
This material is based upon work supported by the National Science Foundation under Grants No. PHY-2317149 and PHY-2110327.

\appendix
\counterwithin{figure}{section}

\section{Plots for Harmonic Trap with \texorpdfstring{$\omega_z = 3\omega$}{b}}

\begin{figure}[ht]
    \centering
    \includegraphics[width=0.95\columnwidth]{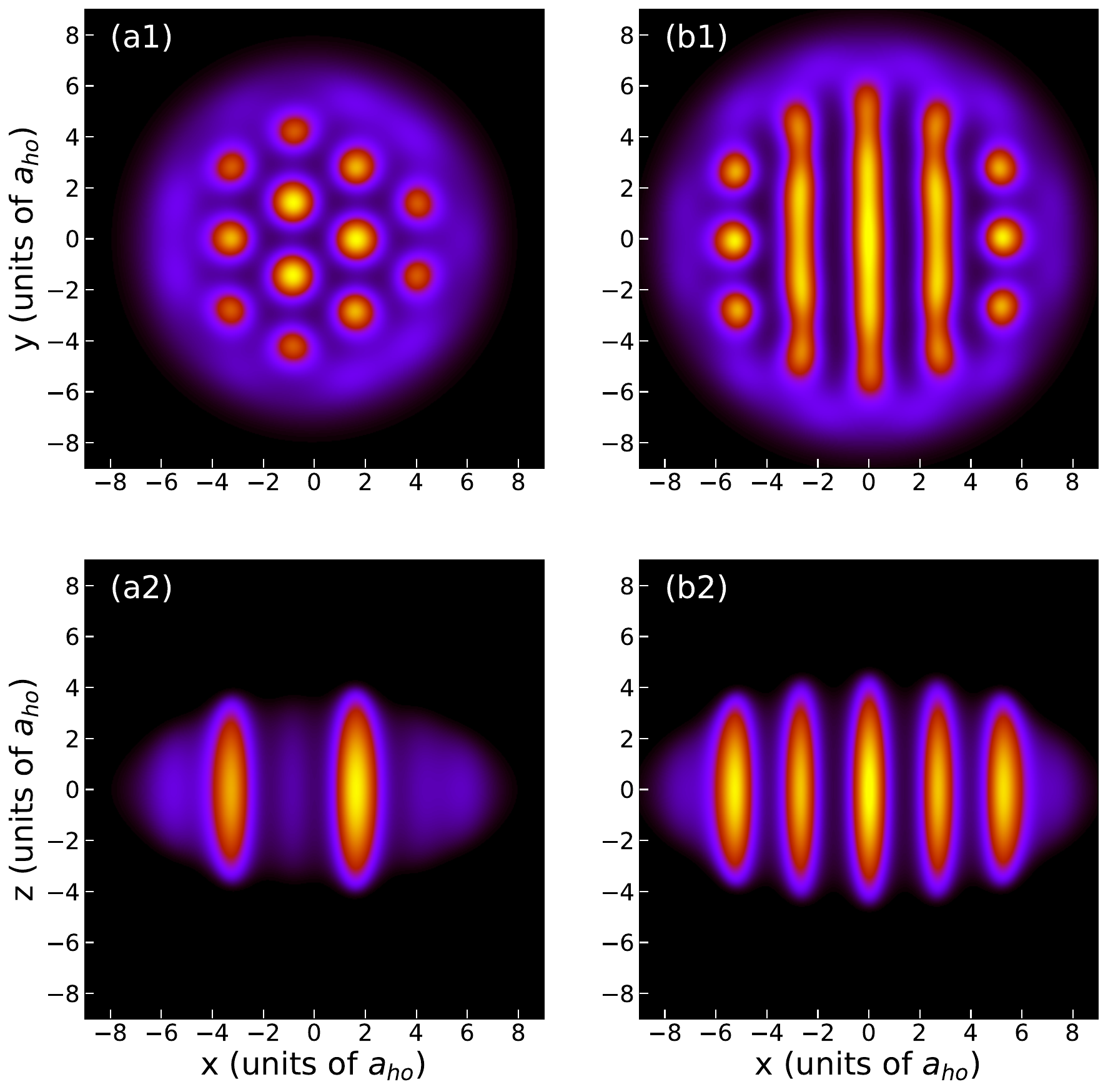}
    \caption{2D Slices of the density profiles of the dipolar BEC of $N=10^5$ (a1), (a2), and $N=2 \times10^5$ (b1) and (b2) $^{162}$Dy atoms confined in the harmonic traps with $\omega_z = 3\omega$.}
    \label{fig:a1}
\end{figure}
\section{Plots for Double Well Trap with \texorpdfstring{$\omega_z = 3\omega$}{b}}

\begin{figure}[ht]
    \centering
    \includegraphics[width=0.95\columnwidth]{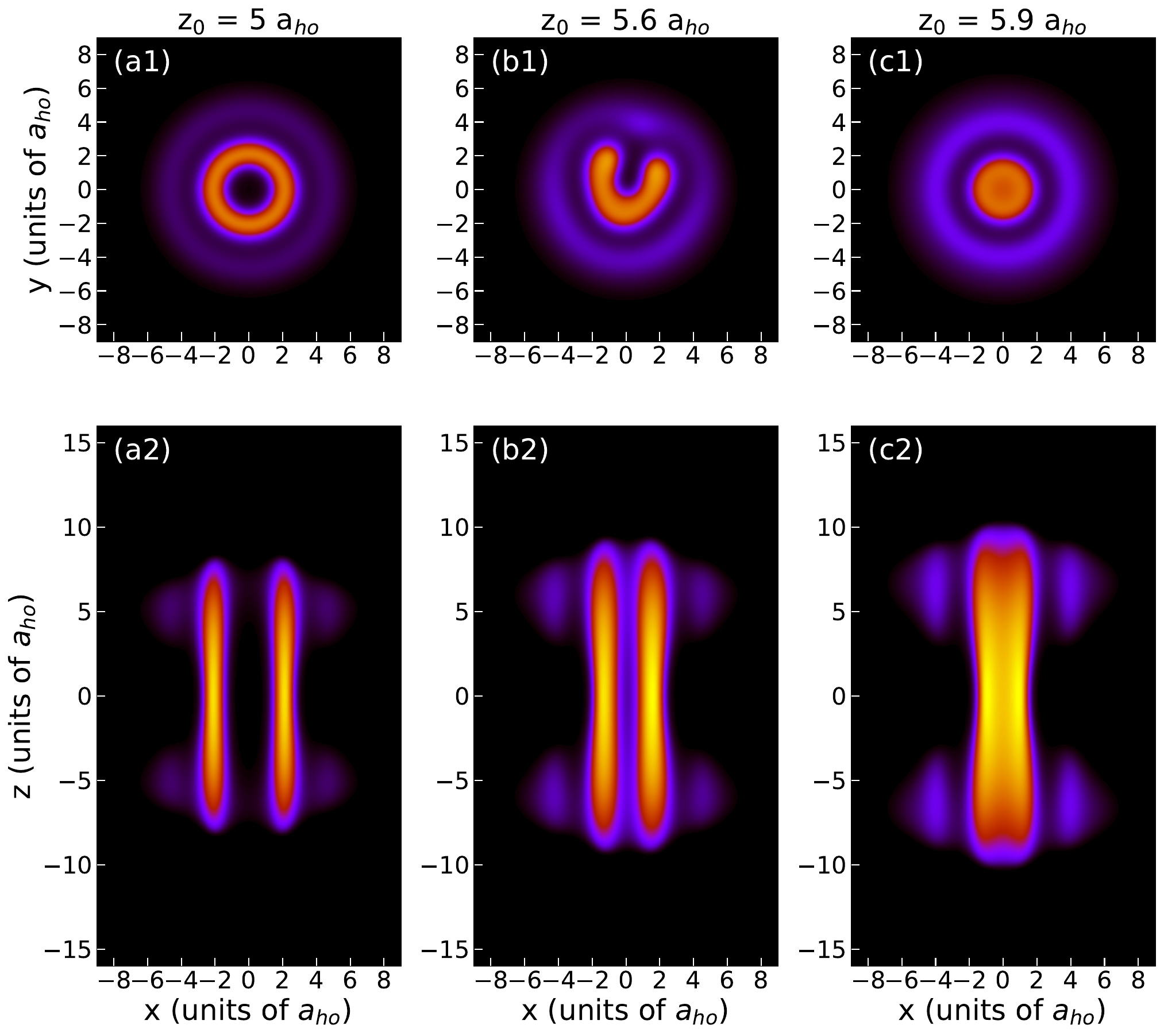}
    \caption{2D Slices of the density profiles of the dipolar BEC of $N=2 \times 10^5$ $^{162}$Dy atoms confined in the double well trap with $\omega_z = 3\omega$. he upper panels show the cut of the $x-y$ plane at the local minima of $z$ with (a1) $z_0 = 5.0 \ a_\mathrm{ho}$, (b1) $z_0 = 5.9 \ a_\mathrm{ho}$, and (c1) $z_0 = 6.6 \ a_\mathrm{ho}$. The lower panels illustrate the corresponding cut along the $x-z$ plane at $y=0$.}
    \label{fig:8}
\end{figure}

\begin{figure*}[ht]
    \centering
    \includegraphics[width=0.95\textwidth]{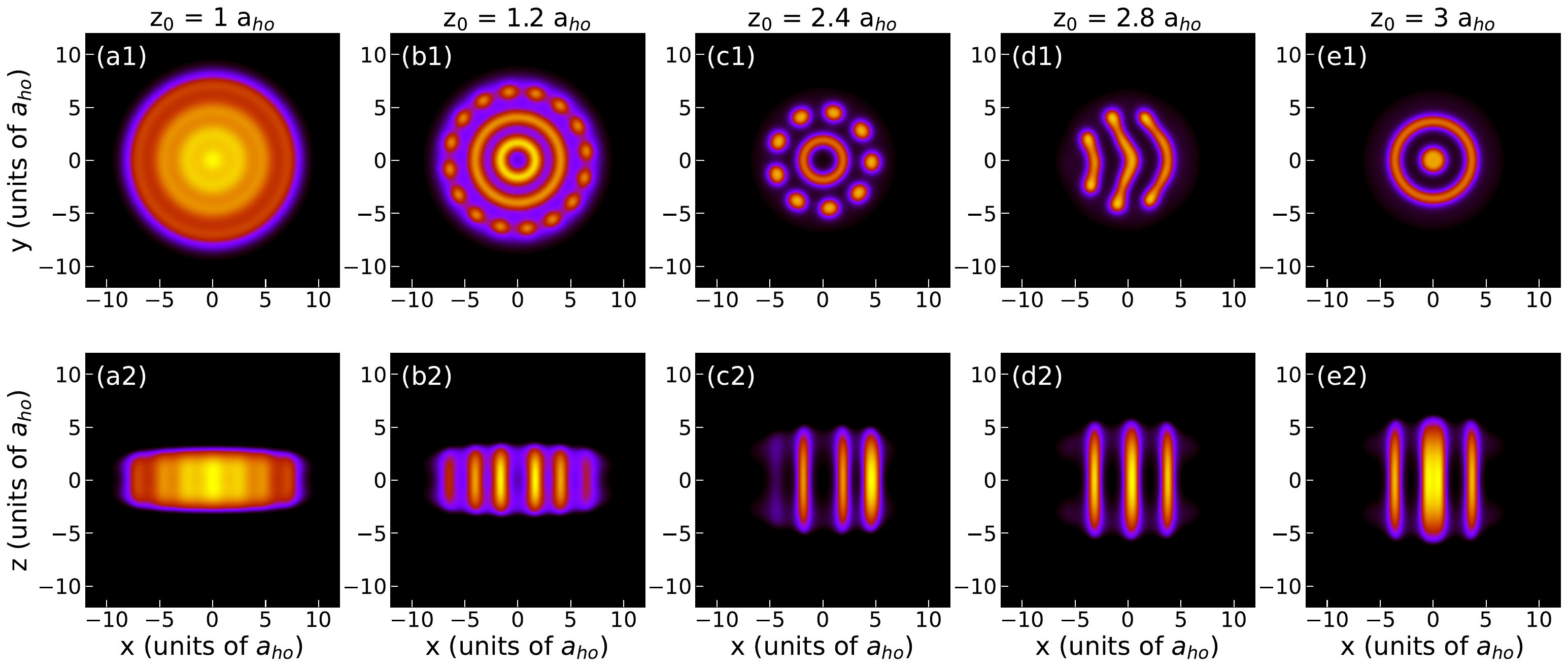}
    \caption{2D Slices of the density profiles of the dipolar BEC of $N=2 \times 10^5$ $^{162}$Dy atoms confined in the double well trap with $\omega_z = 3\omega$. The upper panels show the cut of the $x-y$ plane at the local minima of $z$ with (a1) $z_0 = 1.0 \ a_\mathrm{ho}$, (b1) $z_0 = 1.2 \ a_\mathrm{ho}$, (c1) $z_0 = 2.4 \ a_\mathrm{ho}$, (d1) $z_0 = 2.8 \ a_\mathrm{ho}$, and (e1) $z_0 = 3.0 \ a_\mathrm{ho}$. The lower panels illustrate the corresponding cut along the $x-z$ plane at $y=0$.}
    \label{fig:9}
\end{figure*}

\clearpage
\bibliography{main}% Produces the bibliography via BibTeX.

\end{document}